\documentclass[11pt]{article}
\usepackage[english]{babel}
\usepackage[letterpaper,top=2cm,bottom=2cm,left=3cm,right=3cm,marginparwidth=1.75cm]{geometry}
\usepackage{amsmath}
\usepackage{graphicx}
\usepackage[colorlinks=true, allcolors=blue]{hyperref}
\usepackage[font={small}]{caption}
\usepackage[affil-it]{authblk} 
\usepackage[T1]{fontenc}
\usepackage{etoolbox}
\usepackage{authblk}
\usepackage{cite} 

\begin{document}
\title{\textbf{Fast verification of spent nuclear fuel dry casks using cosmic ray muons: Monte Carlo simulation study}}

\author{A.Sh. Georgadze\thanks{
\href{https://orcid.org/0000-0003-3512-1497}{https://orcid.org/0000-0003-3512-1497},
e-mail: \href{a.sh.georgadze@gmail.com}{a.sh.georgadze@gmail.com},
\href{anzori.heorhadze@ut.ee}{anzori.heorhadze@ut.ee}\\ 
Presented at the 1 st Symposium on Theranostics, 9-11 October 2021, Krakow, Poland}
}
\affil{\footnotesize
$^1$Institute for Nuclear Research of National Academy of Sciences of Ukraine, Kyiv, Ukraine \\
$^2$Institute of Physics, University of Tartu, Estonia}
\date{December 5, 2021}

\maketitle
\begin{abstract}
An application of muon scattering tomography is investigated for the non-destructive verification of spent nuclear fuel dry storage casks in context of international treaty declarations in proving that the diversion of plutonium has not occurred during long term storage in the spent fuel storage facility. It was shown in the previous studies that the method of muon scattering tomography can provide a detailed 3-D image of the spent fuel container but with the measurement time required for that is in the range of a day or even longer. In this paper, we investigate the feasibility of a fast verification of dry storage casks in order to obtain an immediate answer without a detailed 3-D image reconstruction of container content if at least one fuel assembly is missing. We have performed preliminary modelings using Geant4 of safeguards verification of spent fuel dry storage cask HI-STORM 190 with cosmic ray muons. The simulated data were analysed using the Point of Closest Approach method and Kolmogorov–Smirnov statistical tests. To infer the state of content of dry casks we have performed the comparison and further classification of muon observables for the scenarios of one missing fuel assembly with respect to fully loaded casks. The results of this study indicate that one missing fuel assembly can be detected with measurement times of about 1h.

\end{abstract}

\section{Introduction}
With the growing climate change due to carbon emission, nuclear reactors are considered as possible replacements for fossil fuel generation. In contrast to renewable electric power sources which depend on solar luminosity or wind flow, nuclear reactors provide stable electrical power not depending on season or weather changes. Nuclear reactors provide electrical power with no carbon emissions which causes the current global warming crisis. But the increase in nuclear power production is blocked with the problem of permanent safe storage of spent nuclear fuel. Mostly spent nuclear fuel is placed into dry storage casks after several years of cooling down in water pools. Dry casks are steel cylinders that are surrounded by additional steel and concrete to provide neutron and gamma-ray radiation shielding. Dry casks for spent fuel from water reactors typically is a cylinder of diameter   $\approx$ 3 m and height of  $\approx$ 5 m. Each of the casks holds on the average 20-40 spent fuel assemblies.

The plutonium which was produced in fuel rods during the burning uranium process is collected in the spent nuclear fuel and can be chemically separated. The integrity of stored spent fuel should be monitored in the operation of a spent fuel storage facility. All efforts to develop methods based on gamma ray or neutron measurements able to physically determine if the fuel in a cask is in place have failed due to thick shielding provided by the cask walls \cite{Liu1, Greulich}.

At this time there are no methods available for inspecting the internal components of welded stainless-steel canisters once they are loaded. The inspection and verification of spent nuclear fuel stored in dry cask containers is a challenging task which is becoming of increasing importance due to increasing the amount of stored spent nuclear fuel. Therefore, the applicability of muons for nuclear nonproliferation monitoring of spent nuclear fuel dry casks is intensively investigated during recent years \cite{Durham1, Poulson, Durham2,  Park}.

Muons are a natural source of radiation that rain down upon the Earth with an average energy of 5 GeV, at a rate of $\approx$ 10000/m\textsuperscript{2}/minute. As these muons pass through matter they undergo multiple Coulomb scattering. The degree of scatter observed is dependent on the Z of the material. Due to their high penetrating power, muons have the ability to pass through hidden nuclear material and its shielding.
The distribution of zenith angle $\theta$ is known to follow a cosine-squared law, such that $\textit{I}(\theta)$ = $\textit{I}_0 \cos^2\theta$.

Muons like other charged particles lose energy due to ionization. Being minimum ionization particles (MIP), they usually deposit minimal energy while passing through matter. Since muons are charged particles, they undergo multiple Coulomb scattering processes in  matter. The scattering angle distribution depends on the atomic number Z of the material muon is passing and is approximately Gaussian with mean zero and a standard deviation $\sigma$ given by  \cite{gaus,Patr}:
\begin{equation}\label{key}
\sigma = \frac{13.6}{\beta cp}\sqrt{\frac{L}{_{X_{0}}}}(1+0.038)ln\frac{L}{_{X_{0}}})
\end{equation}
\begin{equation}\label{key}
X_{0} = \frac{716.4g/cm^2}{\rho }\frac{A}{Z(Z+1)ln(287/\sqrt{Z})}
\end{equation}
where $\beta$ is the ratio between velocity of muon \emph{v} to velocity of light \emph{c}, \emph{p} is the momentum of the muon, \emph{X\textsubscript{0}} is the radiation length of the material, \emph{L} is the thickness of the scattering
material. \emph{X\textsubscript{0}} is a material property and depends on the density of the material $\rho$, the atomic mass \emph{A} and the atomic number \emph{Z}.

In this work the potential of applicability of muons for monitoring of spent nuclear fuel dry storage casks is investigated.
Muon tomography based on image reconstruction methods can be used to perform scanning of dry casks to verify international treaty declarations in proving that the diversion of plutonium has not occurred including the removal of spent nuclear fuel from dry casks. Usually, to obtain a detailed image of content of dry storage casks more than 40 hours exposure time \cite{Durham2, Braun} is needed with detectors rotating around the cask.

In the works \cite{Chatzidakis,Chatzidakis2} another method was proposed to provide fast and reliable indication of presence of all nuclear materials in the dry cask. The performed in this work Monte Carlo simulations with Geant4 simulation package for 4 GeV monoenergetic muons incident upon a dry cask indicate that for 100000 muons, a few minutes of measurement time with a 3.6 m\textsuperscript{2} detector at a 45° zenith angle, 90\% detection probability for missing fuel assemblies can be achieved with a false alarm rate lower than 1\%. These works explore alternative approaches for in situ monitoring with real-time data analysis and imaging of dry casks using muons in order to provide useful signals about the cask contents in a yes/no decision format much faster than traditional methods based on detailed image reconstruction. Detailed cask imaging can be performed as a second step in the procedure of localization of missing fuel assemblies.

In this work we perform a similar analysis to verify presence of used nuclear fuel in dry cask by applying cosmic ray particle generators CRY \cite{Hagmann} and EcoMuG \cite{ecoMug} coupled with the Geant4 simulation package.

\section{Geometry description and detector layout}
The Centralized Storage Facility for Spent Nuclear Fuel (CSFSF) \cite{Centr} in Ukraine is constructed in the Chernobyl Exclusion Zone. The Holtec International (USA) double-barrier sealing system will be used for  storage of 458 containers type HI-STORM 190 UA with VVER-1000 spent nuclear fuel assemblies stored into multi-purpose containers MPC - 31 type and VVER-440 spent nuclear fuel assemblies stored into multi-purpose containers MPC - 85 type. In this work we have simulated dry cask with MPC - 31 type multi-purpose containers \cite{Report}.

\begin{figure}[h]
	\centering
	\includegraphics[width=.32\textwidth]{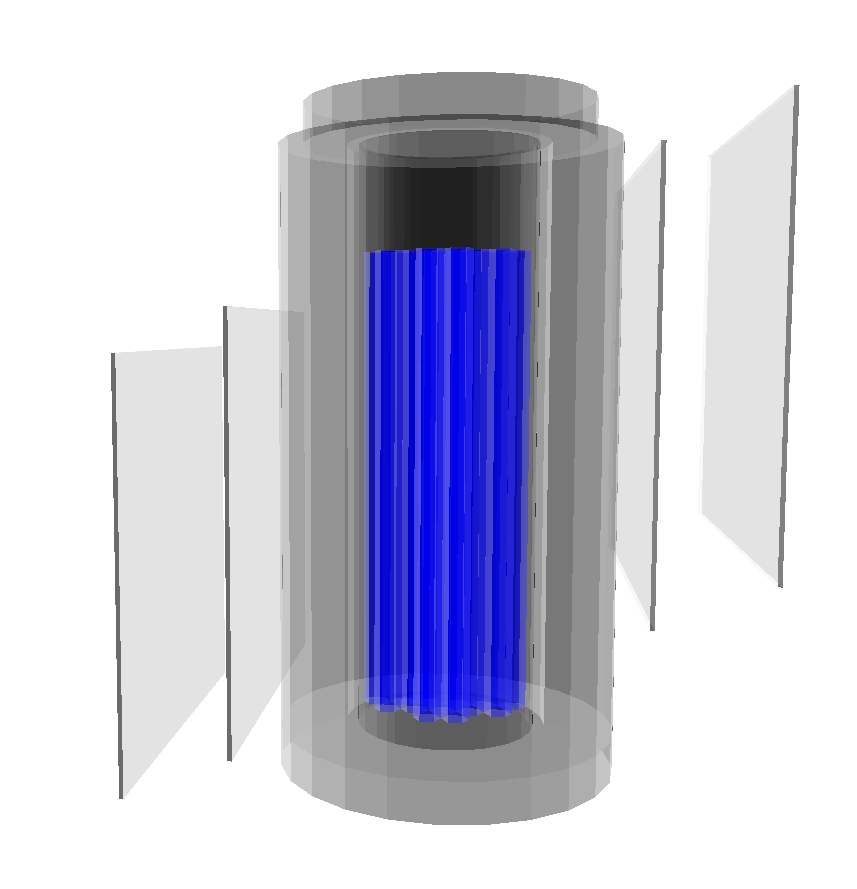}
	\includegraphics[width=.32\textwidth]{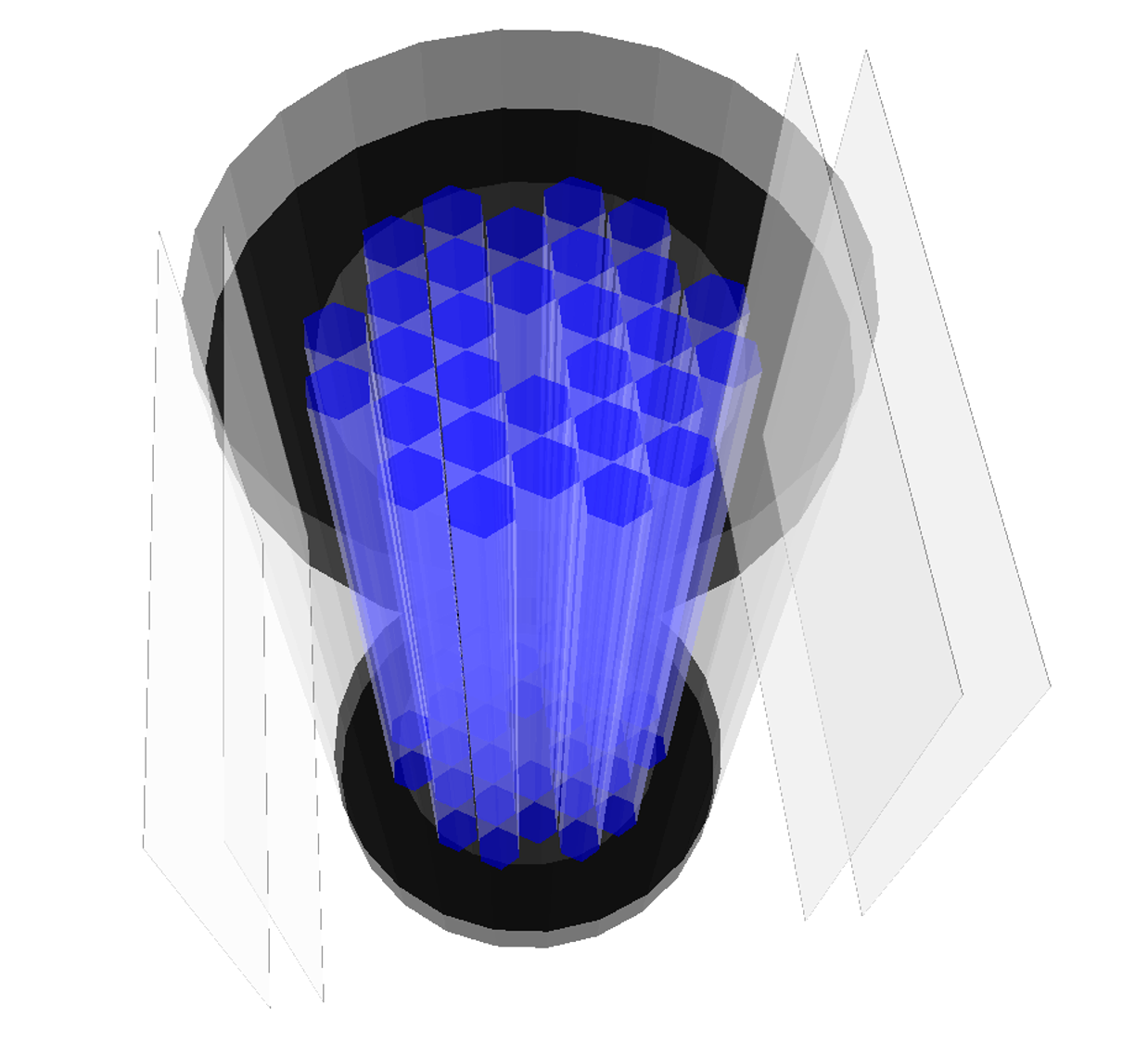}
	\caption{
		Side (left) and top-down (right) illustrations of the dry storage cask with one missing fuel assembly in the center and detectors built in Geant4.}
	\label{Fig:1}
\end{figure}

The geometry of the muon tomography system has been modeled with the GEANT4 simulation package \cite{Agost}. The two pairs of planar tracking detectors are placed on opposite sides of a dry storage cask HI-STORM 190 UA with an MPC-31 type container. This cask holds 31 VVER-1000 spent fuel assemblies. The x- and y- axis are defined as parallel to the ground, while the z-axis is defined as parallel to the normal. The detectors (green color) are modeled as parallel planes with the area of 360 cm × 360 cm with perfect muon hit position information. In order to increase muon flux passing all four detectors and to reduce angular uncertainty detectors are vertically offset against each other as shown in Fig. \ref{Fig:1}. The distance between the pair of tracking detectors is 100 cm. The upper tracking detectors measure  incoming muon track and lower tracker measure hit positions of outgoing muons, which have passed dry casks. Only muons traversing all four planes are tracked and relevant information is recorded into the output file.

\section{Simulation and data sets}
\begin{figure}[h]
	\centering
	\includegraphics[width=1.0\textwidth]{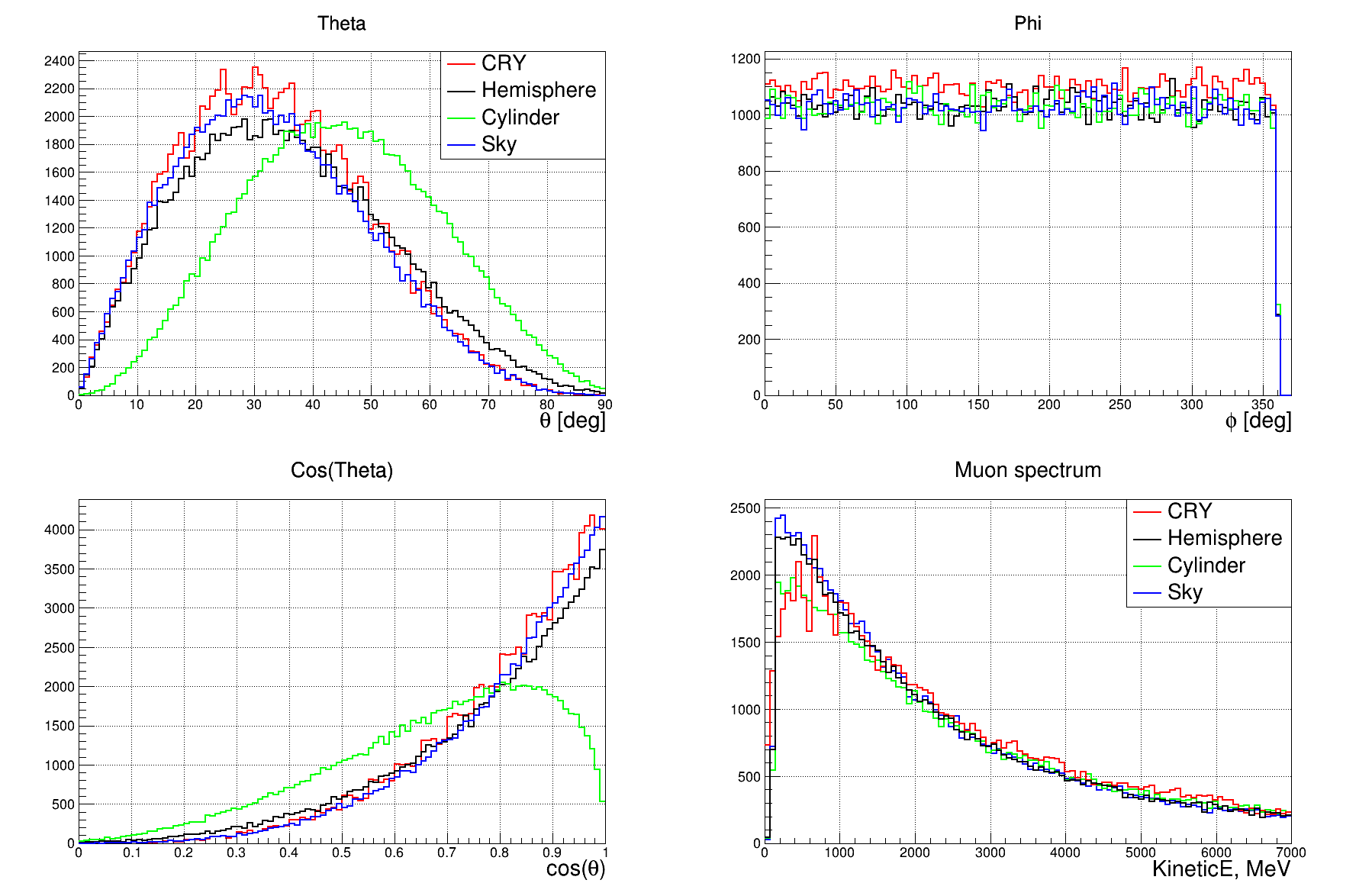}
	\caption{ 
		Comparison between the zenith angle $\theta$ (top left), the azimuthal angle $\phi$ (top right), Cos(Theta) (bottom left) and  kinetic energy (bottom right) for the detected muons for different types of muon sampling surfaces of EcoMug and CRY generators. All spectra are generated for 100000 muons and (20 × 20)$m^2$ generation plane for CRY and EcoMug plane generators and 20 m in diameter for EcoMug cylindrical and a half-spherical surfaces.}
	\label{Fig_cry_ecomug}
\end{figure}

To choose the particle generator we have compared two parametric particle generators - Cosmic Ray Shower Library CRY and EcoMug. Both generators are linked to the Geant4 package and produce particles with the appropriate zenith and azimuth angle distribution of cosmic rays at sea level (see Fig.\ref{Fig_cry_ecomug}). Exception is found for the low energy part of the muon spectrum where there is an excess of muons produced by the EcoMug generator.
In EcoMug, the origin points of generated muons can be sampled from a plane surface (flat sky generation), from a cylindrical surface or from a half-spherical surface. It was found that CRY generators are 3 times faster than EcoMug. But in a real simulation environment this effect is small compared to computational time of muon tracking though the dry cask. The important feature of the CRY package is the presence of an internal subroutine which calculates measurement time for a given number of particles and generation surface. Such a functionality is not present in the EcoMug package.
 
Therefore we have chosen a CRY generator and use it to produce simulated data samples.
The cosmic ray events were generated from a planar surface of 20 m × 20 m area placed 50 cm above the dry storage cask. The Physics List implemented in the simulation accounts for all required physics processes for the transport and interaction of the particles throughout the cask (multiple Coulomb scattering, ionization, bremsstrahlung, pair production, etc.). To emulate realistic detector energy threshold the only muons with deposited energy higher than some settled value were selected. The generated data samples are produced with a measurement time of 30 and 60 minutes. The samples have been produced of two types. The geometry with dry storage casks fully loaded is used to produce the template samples, while the other with one fuel assembly missing is used as a test samples.
Position of the missing fuel assembly was changed to verify the dependence of detection sensitivity to position of empty in different positions inside the MPC - 31.

\section{Muon reconstruction}
We have used a muon tomography reconstruction algorithm based on the Point of Closest Approach (POCA) method \cite{Sunday, MuonPortal} which calculates the closest point of approach between two 3D lines. The POCA algorithm makes the simplified assumption that the muon scattering occurs in a single-point and searches for the point of closest approach between the incident $\textit{v}_{in}$ and outgoing $\textit{v}_{out}$ reconstructed muon track directions with respect to the dry storage cask. In 3D the two tracks may not intersect and the shortest line segment between the tracks is estimated by finding the pair of points of closest approach between the two lines. The midpoint of this line segment is considered as a scattering point of the muon, so called POCA point. The recorded data on hit position on each detector layer are used to produce two tracks constructed using hits produced by muon in the top and bottom tracker detectors, to calculate POCA point and the scattering angle between the two tracks by the use of formula \cite{Poulson, Carlisle}:
\begin{equation}
\theta_{scatt} = \arccos(\frac{\vec{v}_{in} \times \vec{v}_{out}}{\left | \upsilon_1 \right |\left | {\upsilon_2} \right | })
\end{equation}

The realistic hit position resolution of muon trackers is taken into account by reconstructing the POCA points with hit positions in the tracking detector smeared simultaneously in x and y directions by Gaussian with resolution $\sigma$ of 0.15 mm.
This resolution corresponds to detector design based on multiple layers of scintillating plastic fiber arrays used as position sensitive detectors.
The prototype based on such a design was constructed \cite{Anbarjafari} and spatial resolution of 0.15 mm was obtained with a double layer array of round cross-sections plastic scintillating fibers of 1 mm in diameter.

\section{Results}
\begin{figure}[h]
	\centering
	\includegraphics[width=\textwidth]{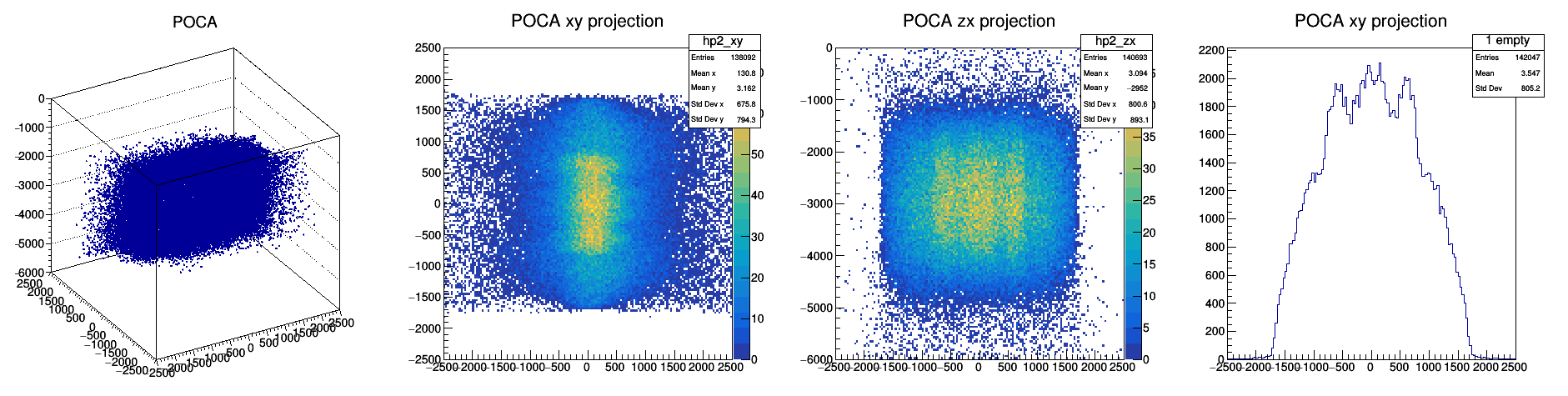}
	\caption{Left: POCA cloud in 3-D view. Middle left and middle right: POCA cloud 2-D projections on the x-y and x-z planes, right: 1-D projection of POCA cloud. POCA points are reconstructed for scattering angles greater than 50 mrad.}
	\label{Fig_poca1}
\end{figure}

\begin{figure}[h]
	\centering
	\includegraphics[width=\textwidth]{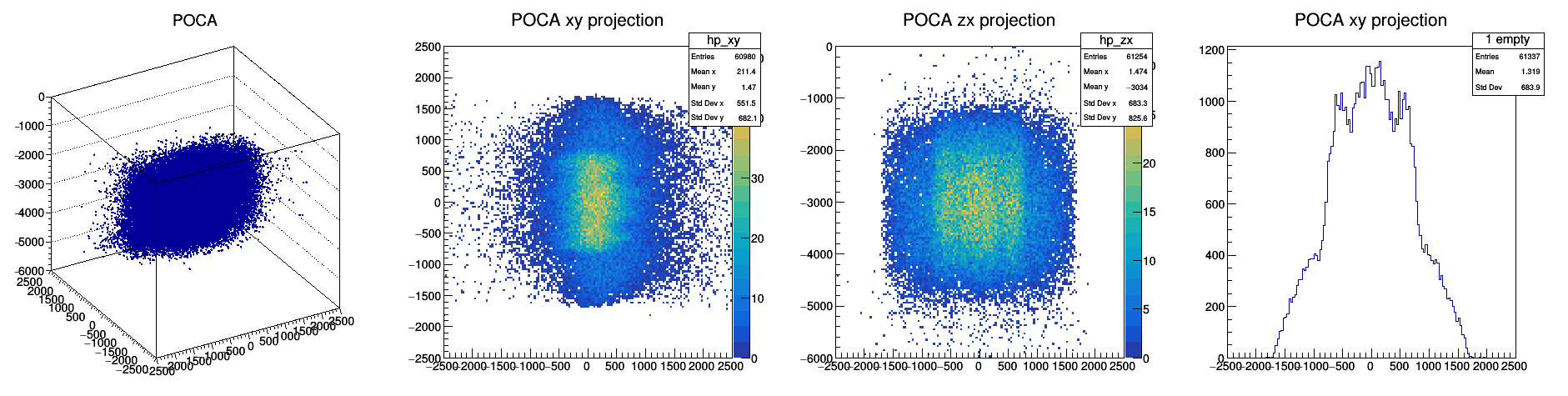}
	\caption{Same histograms as previous ﬁgure but with POCA points reconstructed for scattering angles greater than 100 mrad.}
	\label{Fig_poca2}
\end{figure}

The simulated template and testing data samples contains the three-dimensional spatial distribution of the scattering point with a weight proportional to the value of the scattering angle. Both types of data samples are analyzed for consistency using Kolmogorov-Smirnov test using similar approach published in \cite{arbol}. One missing fuel assembly in the dry cask resulted in distortion of POCA density which is abnormal compared to POCA density of fully loaded dry cask. Using the ROOT data analysis package \cite{ROOT} we have performed simple Kolmogorov tests of POCA density data of generated data samples. It was found that best sensitivity was obtained performing K-test of the two-dimensional x-y projections of POCA density which correspond to a top view on a dry cask container. To remove noise contamination from scatterings in the air and low-Z materials we have used in the analysis reconstructed POCA points on muon tracks which scatter with an angle greater than 50 mrad. The resulting three-dimensional distributions of scattering points and their 2-D and 1-D projections are drawn in Fig. \ref{Fig_poca1} and Fig. \ref{Fig_poca2} for 60 minutes measurement time and scattering angles larger than 50 mrad and 100 mrad correspondingly. 
A first look at the POCA results reveals that since POCA method neglects the multiple scattering processes this resulted in a smeared dry cask image. Nevertheless, the outer and inner shapes of the dry cask can be recognized.  
Even if the reconstructed images of dry storage casks are not precise, the distributions are different enough to perform statistical tests to check the compatibility of template samples against test samples. In more detail, we compare simulated template samples against each other and compare simulated test samples against template samples. 

\begin{figure}[h]
	\centering
	\includegraphics[width=.32\textwidth]{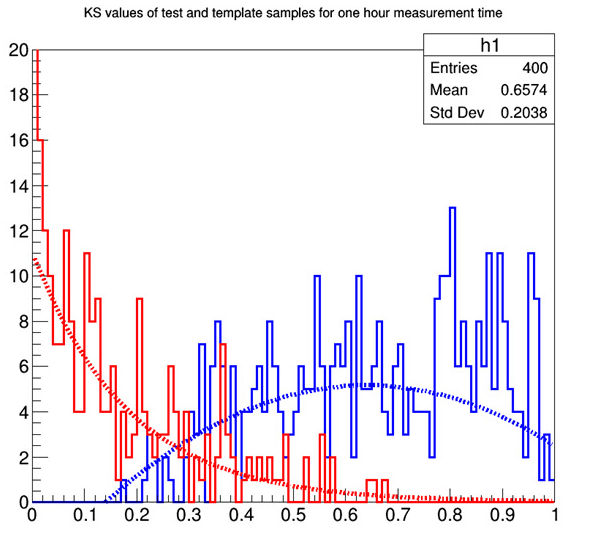}
	\includegraphics[width=.34\textwidth]{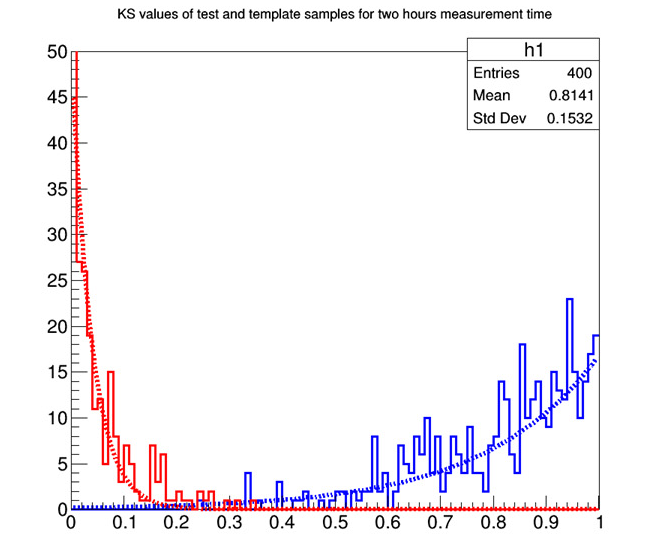}
	\includegraphics[width=.32\textwidth]{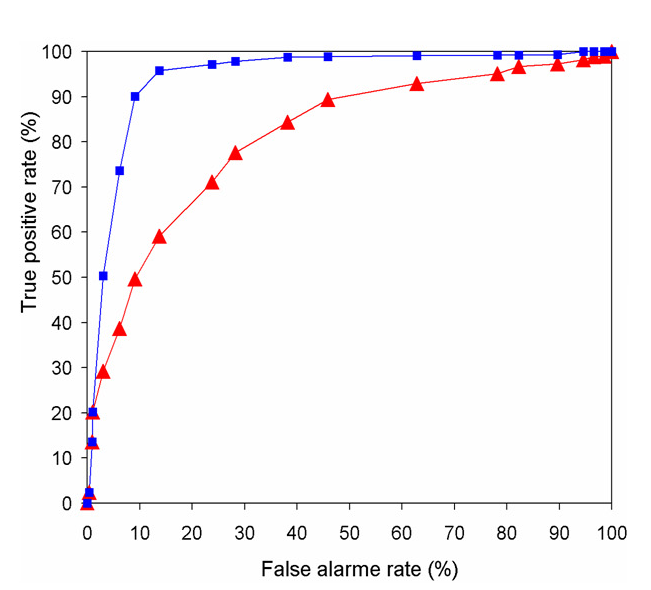}
	\caption{\textit{Left and middle panels}: The scores obtained for K-test values calculated for test samples (red color) and template samples (blue) for 30 minutes 60 minutes measurement times correspondingly.
		Right: ROC curves for or 30 minutes (red) and 60 minutes (blue) measurement times. \textit{Right panel}: ROC curves for or 30 minutes (red) and 60 minutes (blue) measurement times.}
	\label{Fig_KS}
\end{figure}

Figure \ref{Fig_KS} (left and middle) represent histograms with plotted scores obtained with the Kolmogorov-Smirnov test calculated for all of the template and test data samples for two  measurement times. 
We have used two-sample Kolmogorov-Smirnov test to verify whether template samples and test samples come from the same distribution. The null hypothesis is H\textsubscript{0}: both simulated data samples come from a population with the same distribution. 
Without statistical fluctuations in POCA point distribution two-sample Kolmogorov tests of template samples against template samples must be equal to 1 since they are result of measurements of fully loaded casks, that is comes from same distribution. Due to fluctuations K-test values for template samples form the distribution which is grouped close to 1 if fluctuations are smaller for longer measurement time and broad distribution for shorter measurement time.  
Alternatively, applying two-sample Kolmogorov-Smirnov test to classify template samples against test samples K-test value must be equal to 0 because they come from a populations with the different distributions. Due to fluctuations K-test value are grouped close to 0 in case of small fluctuations for 60 min measurement time and form broader distribution for 30 minutes measurement time which cause intersecting K-test values for template and test samples. 
One can see that for the 30 minute measurement time due to smaller number of muons used for reconstruction fluctuations are larger resulting in considerable intersection of Kolmogorov test values for template and test data samples. 
For the 60 minute measurement time number of muons used for reconstruction is twice larger and K-test values became well separated. 
In order to study quantitatively the ability of muon tomography system to detect missing nuclear fuel assembly in dry storage cask the Receiver-Operator Characteristic (ROC) curve \cite{Obuchowski,Fawcett} was generated for measurement times of 30 minutes and 60 minutes. This graph characterizes the trade-off between the true positive rate of correctly identifying missing fuel assembly as a function of the false positive rate of incorrectly identifying fully loaded dry casks as  one with missing fuel assembly. ROC curves allow quantifying the performance of applied Kolmogorov classifiers. A possible value for the quality of the discrimination performance is the area under the curve (AUC) of the ROC curve. The AUC was calculated in a percentage and 100\% means an ideal classification with total separation between the two scenarios of fully loaded casks and one with missing fuel assembly.

On Fig. \ref{Fig_KS} (left) one can see that for 30 minutes measurement time there is a considerable overlap of K-test values of template and test samples. The corresponding AUC value is $\approx 80\%$, that means there is a high probability to mistakenly identify a cask with missing fuel assembly as a fully loaded cask.
Different cut values will result in different probabilities of incorrect identification but for high efficiency of discrimination performance more data and consequently longer measurement time is needed.
K-test values for template and test samples for 60 minutes measurement presented on Fig. \ref{Fig_KS} (middle). Their overlapping is considerably smaller and corresponding AUC is $\approx 96\%$ is close to perfect classification.
These results correspond to a conclusion drawn in \cite{Chatzidakis} that the scattering distributions between a fully loaded dry cask and one with a fuel assembly missing overlap significantly for small number of muons passing cask during short measurement time but their discrepancy eventually increases with increasing number of muons.
It should be noted that more sophisticated methods based on Machine Learning can be applied to improve detection efficiency and decrease measurement time.

\section{Discussion}
A simple detector set-up composed of four tracking detectors has been considered to simulate muon transport through the dry casks and reconstruct POCA points. In this preliminary study we have applied CRY cosmic ray particle generator for realistic muon flux production and have shown that without detailed imaging of dry casks using the POCA method and statistical analysis one missing fuel assembly can be detected with relatively short measurement time of one hour with detection efficiency  $\approx$ 96\%. This time is longer than that estimated in \cite{Chatzidakis,Chatzidakis2} but this would be a result of applying simplified method of POCA reconstruction which is purely geometrical and as a result produce a lot of POCA points lying outside the detector area which results in a smeared image and decreased detection efficiency. More sophisticated reconstruction methods will be applied to obtain more precise characterization of muon tomography set up.
In the real environment in case of detection of missing fuel assembly with a proposed statistical method the detailed image can be acquired with longer measurement time performing rotation of detector around the cask. Then  location of missing assembly can be found in the way as it is described in \cite{Zhengzhi}.

The results of this work demonstrate that using scattered muon tomography makes possible prompt monitoring of spent nuclear fuel dry storage casks for nuclear non-proliferation purposes which is not possible with gamma or neutron scanning and imaging techniques due to too heavy shielding of container with used nuclear fuel.


\end{document}